# Neural Networks for Modeling and Control of Particle Accelerators

A. L. Edelen, *IEEE Student Member,* S.G. Biedron, *IEEE Senior Member,* B.E. Chase, *IEEE Member,*
D. Edstrom Jr., S.V. Milton, *IEEE Senior Member*, P. Stabile, *IEEE Member*

*Abstract*—Particle accelerators are host to myriad nonlinear and complex physical phenomena. They often involve a multitude of interacting systems, are subject to tight performance demands, and should be able to run for extended periods of time with minimal interruptions. Often times, traditional control techniques cannot fully meet these requirements. One promising avenue is to introduce machine learning and sophisticated control techniques inspired by artificial intelligence, particularly in light of recent theoretical and practical advances in these fields. Within machine learning and artificial intelligence, neural networks are particularly well-suited to modeling, control, and diagnostic analysis of complex, nonlinear, and time-varying systems, as well as systems with large parameter spaces. Consequently, the use of neural network-based modeling and control techniques could be of significant benefit to particle accelerators. For the same reasons, particle accelerators are also ideal test-beds for these techniques. Many early attempts to apply neural networks to particle accelerators yielded mixed results, due to the relative immaturity of the technology for such tasks. The purpose of this paper is to re-introduce neural networks to the particle accelerator community and report on some work in neural network control that is being conducted as part of a dedicated collaboration between Fermilab and Colorado State University (CSU). We describe some of the challenges of particle accelerator control, highlight recent advances in neural network techniques, discuss some promising avenues for incorporating neural networks into particle accelerator control systems, and describe a neural network-based control system that is being developed for resonance control of an RF electron gun at the Fermilab Accelerator Science and Technology (FAST) facility, including initial experimental results from a benchmark controller.

A.L. Edelen, S.G. Biedron, and S.V. Milton are with the Department of Electrical and Computer Engineering, Colorado State University, Fort Collins, CO 80523 (email: auralee.morin@colostate.edu). S.G. Biedron is also with the Department of Electrical and Computer Engineering, University of Ljubljana, Slovenia, Tržaška 25, SI-1000 Ljubljana, Slovenia.
B.E. Chase and D. Edstrom Jr. are with Fermi National Accelerator Laboratory, Batavia, IL 60510-5011
P. Stabile is with ADAM (CERN spin off), Geneva, Switzerland and was formerly with Fermi National Accelerator Laboratory, Batavia, IL 60510-5011.

## I. INTRODUCTION

Particle accelerators are host to myriad complex and nonlinear physical phenomena. Adding to this inherent complexity, they often involve a multitude of interacting systems, exhibit long-term process cycles, and endure changes in individual machine components over time. In addition, they are often subject to tight tolerances on beam parameters and other performance metrics, and it is often desirable for them to run for extended periods of time with minimal interruptions. In addition, many particle accelerators are concurrently supporting a variety of request-driven or cyclic processes (i.e. they are often not running in a steady-state condition). There will also inevitably be deviations between the system design, numerical or analytic physics-based simulation models, and the installed system. Finally, as increasingly high-intensity, high-energy, and high-gradient accelerators are built that fundamentally rely on increasingly complex/nonlinear phenomena, traditional control techniques become inadequate in some domains. Taken together, this leaves us with many challenges for designing control systems that will reliably meet performance demands for both present and future accelerators.

These challenges can become more acute for applications of particle accelerators in medicine, industry, and defense. These applications range from relatively well-established use cases where increased automation and better control could be of significant benefit (e.g. particle beam therapy for cancer treatment), to as-yet unrealized applications that require substantial improvements in controller robustness, flexibility, and/or portability before they will be feasible (e.g. compact, high-average-power FELs for EUV lithography). Furthermore, outside of large accelerator facilities, day-to-day reliance on highly-skilled operators and technicians is often undesirable.

One avenue toward meeting these challenges is the incorporation of recently improved techniques from the fields of machine learning (ML) and artificial intelligence (AI) into the design of control systems for particle accelerators. In particular, techniques based on neural networks (NNs) are well-suited to modeling, control, and diagnostic analysis of complex, time-varying systems, and systems with large parameter spaces [1, 2]. These techniques can be used in conjunction with actual machine data, thereby accounting for noise, variable delays, subtle statistical correlations, and complex effects that may not be easily addressed *a priori*. NNs can also be useful in cases where accurate data from simulations or some other computationally intensive procedure is available, but the input-output relationship needs to be computed more rapidly for effective real-time deployment. Because of their functional flexibility, they are able to operate effectively for many different kinds of tasks.

Here we present an overview of some challenges in particle accelerator control, provide an overview of relevant AI concepts, describe some ways in which we are applying these to accelerators, and present an example of our work at FAST.

The remaining discussion is organized in the following manner. Sections *II.A—II.C* provide an overview of some challenges encountered in particle accelerator control and collectively provide the corresponding motivation for developing AI- and ML-based control techniques for particle accelerators. Section *II.D* highlights some of the ways in which the skills employed by accelerator operators can be used to inform the design of AI-based control schemes. Section *II.E* briefly describes several other advanced methods that the particle accelerator community is pursuing at present to improve performance. Section *III* provides some definitions and very basic technical background on ML, AI, and NNs. Section *IV.A* highlights some proposed use cases for NNs in the modeling, control, and diagnostic analysis of particle accelerators, many of which the authors are presently pursuing or plan to pursue. Section *IV.B.1* describes some of the developments that have dramatically improved the practical usefulness of NNs in recent years. Section *IV.B.2* highlights some previous efforts (both successful and unsuccessful) to apply NNs and AI to particle accelerators. Section *IV.B.3* provides some examples of recent successes in other scientific and engineering disciplines. Finally, section V describes some initial results and planned work for NN-based resonance control of an RF gun at FAST.

## II. CHALLENGES FOR PARTICLE ACCELERATOR CONTROL AND MOTIVATION FOR THE USE OF MACHINE LEARNING AND ARTIFICIAL INTELLIGENCE

### A. Preliminary Terminology

In the following discussion, we will use the term "optimization" to indicate an iterative search process through which better combinations of operating parameters are found such that specific performance goals are better met. This could in principle be done manually by a human operator or automatically using a mathematical optimization routine.

We will use "control" to indicate a dedicated process encoded by a set of rules through which a set point, series of set points, or other set of performance goals is achieved and maintained despite the presence of disturbances in the machine. "The machine," for our purposes refers to a particle accelerator system or sub-system.

We will use the term "tuning" as shorthand for one kind of task that a human operator performs: adjusting a setting or group of settings such that criteria for good performance are met. Here, tuning is not strictly equivalent to optimization in the sense described above (i.e. search). Rather, tuning requires the operator to combine elements of model learning, control policy learning, planning, and prediction as well.

Finally, the terms "online" and "offline" are used differently in different disciplines.[1] Here, we will follow typical use within the accelerator community: "online" will indicate that a given computational procedure is running and interacting with the machine concurrent to operation, and "offline" will indicate a process that does not run in this manner (e.g. using a simulation to find optimal parameter settings prior to running the machine, or analysis of data gathered from a diagnostic after an operating run has been finished).

### B. Challenges for Human Operators

Typically, particle accelerator systems and subsystems are extensively simulated and optimized. Once running, data from the machine can be used to update these offline models and provide a more accurate set of predicted optimal settings. Even with such measures, operators will often conduct extensive tuning each time the machine is put into a new operating condition or turned on after a shut-down. This can work well for dynamics that operate on a few human-compatible timescales (i.e. ones that are not too long or too short—hundreds of milliseconds to tens of minutes) or that involve only a few parameters. However, the task can become unwieldy as the dynamics become more nonlinear or as the number of parameters and interrelations increases. The presence of multiple timescales of behavior can also make isolation of relevant parameters extremely difficult.

Experienced accelerator operators can become adept at handling complex dynamics and many parameters quickly, particularly on machines with which they are very familiar and/or for tasks that are frequently repeated. However, clearly there is a point at which even the most capable human will not be able to efficiently and effectively synthesize all of the information required to achieve good performance. Furthermore, machines that require frequent changes in beam parameters or operating conditions vastly increase the number of learned control strategies and specific procedures that operators must employ.

Ultimately, human operation is limited in the following ways:
1) Humans can only process a handful of input parameters at once;
2) Humans can only act on a few parameters at once;
3) Humans can only operate on a relatively narrow set of timescales, and separating multiple timescales during problem diagnosis can rapidly become infeasible as the number of these increases;
4) Humans are expensive and their skill levels may vary dramatically even in the execution of a standard procedure.

Instead, one would like to automate many of the routine tasks historically handled by operators (preferably in a way that does not itself require extensive, ongoing human intervention to function properly). Time lost during tasks such as tuning tends to be expensive, both with regard to the personnel and energy costs incurred by running the machine and with regard to the scientific goals of researchers who are generally allocated a limited amount of machine time in which to run an experiment. As such, sufficiently automating routine tasks can significantly improve performance in terms of the beam parameters achieved, time efficiency, and overall operational costs (both in the narrow sense of cost per unit time and in the broader sense of cost per experimental investigation).

---

[1] For example, in neural network training, "online" can mean incremental training as samples become available and "offline" can mean batch training with a segment of previous samples—which can still occur concurrently with the process that is producing training samples.

### C. Challenges for Automated Systems

Despite the limitations noted above, humans are remarkably good at collecting disparate kinds of information and putting them to judicious collective use in ways that are challenging for many automated systems and traditional control techniques at present.

For example, some machines have limited diagnostics, and thus one must properly adjust a large number of variables using just a few measureable outputs. Without the human-level knowledge and deductive reasoning ability of the operator, it can be very difficult to reliably automate this kind of process. Conversely, some machines have many diagnostics that ideally should be used both individually and collectively (e.g. by extracting higher-level state information from a variety of readings and generating an appropriate response). Nuances in the way the evaluation is made in an automated system can result in poor decisions that the designers of the system would not have anticipated and that a human operator would have rightly never considered.

Another example centers around planning. Systems that involve significant time delays relative to the timescale on which adjustments must occur may also benefit from incorporating planning what the next series of actions should be given some predicted and desired behavior. That decision process itself can be difficult to automate, let alone codifying the required representations of the system dynamics in a way that is sufficiently accurate and can be executed quickly. The complexity of accelerators increases this difficulty substantially. Thus, a process that is relatively simple for an operator becomes a challenging task for an automated system.

In taking stock of some tasks and capabilities that would ideally be achieved with an automated system, the challenges become more apparent. We may wish for such systems to be able to do the following:
—Make efficient use of high-fidelity models for online use in control routines (e.g. for prediction or for filling in details on behavior for which diagnostics are unavailable);
—Create models that can be continuously adapted to match the real machine;
—Identify and compensate for long-term process cycles and drift (due to hardware aging, incremental component replacements, and slowly varying dynamics that are not accounted for in other ways);
—Compensate for deviations from the ideal design, such as noise, misalignment, and deleterious effects arising from system interactions;
—Quickly distill large amounts of data into useful information, even for cases where data analysis is not straightforward, so that it can be used effectively in a control system or by an operator—this is applies to both system-wide higher-level diagnostic analysis and readings from individual components;
—Simultaneously optimize machine parameters system-wide to maximize overall performance metrics, as optimizing just one subsystem or set of parameters may produce undesirable results overall;
—Perform rapid adjustment of settings in the face of new operating conditions (e.g. new beam parameters);
—Take pre-emptive control actions where necessary and find a good series of future control actions to achieve a desired set of predicted outputs;
—Strictly adhere to hard constraints and allow reasonable violation of soft constraints.

### D. Expanding the Scope: Artificial Intelligence and Neural Networks

When trying to achieve human-level performance in a given task, it is useful to think about what the human operator does in that context. Often, significant advances can be made by critically examining the real-world human problem solving process and breaking it into constituent parts for a given task. Take, for example, a recent advance in NN-based recognition of written characters that significantly outperforms competing methods, even with far fewer training examples: the key insight was to incorporate a process through which the NN learns to mimic the way a human learns to *produce* letters in the first place, i.e. "learning to learn," rather than simply providing extensive training on a large data set of examples [3]. Thus, a human-inspired process that at first might seem tangential to a typical character recognition task was in fact of substantial importance for achieving improved performance.

In the context of accelerator operators, we see first that the operator has an understanding of the dynamics of the machine (i.e. a system model) through both a theoretical understanding of the ideal behavior of the machine and through the observed behavior of the machine. This model is adapted through experience and can be compared with similar machines for further insights.

Second, the operator has and uses memories of previous states visited, actions taken, and resultant outcomes to help them form better control policies and return to previously visited operational states efficiently.

Third, the operator has the ability to plan a series of future control actions based on their mental model of the machine.

Fourth, the operator can do fast data reduction, image processing (e.g. taking visual input from a diagnostic and making inferences based on it), and pattern recognition (e.g. recognizing when an instability is starting to develop).

Finally, an operator with many years of experience or deep knowledge of a given system does not typically need to solve partial differential equations or run a physics-based simulation to have a good idea of what will happen when they take certain actions in certain machine states, even when the outcome of a specific combination of states and actions has not previously been observed—they have learned fast, heuristic representations of the relevant processes that allow for generalization beyond direct experience and memory.

Thus, operators are not merely searching for the best combination of machine settings that produce the desired beam parameters at any given time (as an online stochastic optimization procedure applied directly to controllable parameters would). They are also making predictions based on mental models of the system, checking observed behavior against the models to improve them over time, planning series of future control actions, interpreting a substantial amount of diagnostic information, and using present and past performance to adjust the rules by which control decisions are made in various sets of observed or inferred machine states (i.e. developing and remembering control policies). In short, operators simultaneously use a combination of optimization, model learning, planning, prediction, diagnostic analysis, and policy learning to control the machine.

For each of these capabilities of an experienced operator, there is an analogous set of techniques in ML, AI, and advanced control. If we can capture some of these capabilities while also circumventing some of the limitations of manual operation, we can address the control challenges described earlier far more effectively. Our work in developing NN-based control systems for particle accelerators is specifically guided by this line of thinking.

*E. Some Other Recent Approaches Toward Achieving Greater Automation and Improved Performance*

Here, we will very briefly highlight some other advanced modeling, optimization, and control approaches being pursued within the accelerator community.

First, online optimization using stochastic optimization methods is being pursued in various forms (see, for example [4, 5]). Techniques such as particle swarm optimization (PSO) and genetic algorithms (GAs) have been of particular of interest [6, 7].

A promising technique based on extremum-seeking control has recently been developed [8, 9], and it has been used for control, optimization, and prediction of multiple parameters in several particle accelerator applications [10, 11, 12, 13].

Finally, there have been many recent advances in online modeling for particle accelerators that significantly improve the computation speed of high-fidelity physics-based models. This can be accomplished through both parallelization (e.g. GPU acceleration) and establishing a judicious balance in the tradeoff between the speed and accuracy of the calculations employed. A pioneering example is given in [14].

III. ARTIFICIAL INTELLIGENCE, MACHINE LEARNING, AND NEURAL NETWORKS: BACKGROUND AND DEFINITIONS

*A. Machine Learning and Artificial Intelligence*

Broadly speaking, *machine learning* is concerned with improving the performance of an algorithm on some task over time through interaction with data (e.g. learning to make predictions about a system, learning to recognize handwritten characters, learning to detect aberrant credit card activity). It is a sub-field of *artificial intelligence*, which is concerned more generally with creating systems that are capable of behaving "intelligently" (i.e. creating intelligent agents). Though the exact definitions used within the field vary, behavior is generally considered to be "intelligent" when it includes some combination of planning, interpreting environmental input, self-assessment, adaptation of behavior in response to the environment, and rational decision making.

ML and AI rely heavily on techniques from computational statistics and stochastic optimization.[2] Indeed, the lines between these fields are also very blurry, particularly as methods become increasingly hybridized.[3] ML techniques also tend to become absorbed by relevant surrounding fields once they are well-established.[4]

Some typical tasks in ML include *classification* (categorizing instances of data), *clustering* (collecting similar kinds of data together), *dimensional reduction* (reducing the number of random variables by finding and exploiting relationships between them or mapping them to a *new* set of variables), and *regression* (estimating relationships between variables).

Some typical frameworks for learning include *supervised learning*—in which examples demonstrating correct relationships are given (e.g. data with labels, such as pairs of input-output data), *unsupervised learning*—in which no specifically correct examples are given and underlying structures in the data must be found on their own (e.g. unlabeled data that must be grouped into similar categories without specifying what those categories are), and *reinforcement learning*—in which an agent interacts with the environment and alters its behavior based on the "reward" it receives. Methods which combine these frameworks, as well as more specialized learning paradigms,[5] are becoming increasingly common.

*B. Reinforcement Learning*

Supervised and unsupervised learning are fairly intuitive concepts, and as such it is relatively straightforward to understand how one might go about implementing them. Reinforcement learning (RL), however, requires a little more description. In a RL task, an agent learns how to respond to its environment (i.e. it learns a *policy*) such that some representation of its performance is maximized over time.

A traditional reinforcement learning scheme typically includes the following components:
1) A *policy* which maps observed system states to actions (i.e. these are rules by which control actions are chosen);
2) A *reward function* that delivers a scalar value indicating how "good" the environmental response to the chosen behavior is;

---

[2] For a simple example, consider that the connections in a neural network can be trained using gradient descent, genetic algorithms, PSO, or any other standard stochastic optimization technique, and improvements in these optimization techniques open the doors for increasingly complex neural network structures to be trained efficiently and reliably.

[3] Take, for example, reactive search optimization (RSO) [15,16], which uses machine learning to automate the choice of algorithm parameters in more traditional optimization methods.

[4] As one well-known researcher, John McCarthy, purportedly lamented "as soon as it works, no one calls it AI anymore"—a sentiment echoed by many in the field.

[5] For example, take *transfer learning*, which specifically focuses on how to reliably transfer previously learned information to new situations or problem domains. Similarly, *learning to learn* is inspired by the way humans optimize the way they learn to complete new tasks through experience.

3) A means of estimating long-term future expected rewards for given states or state-action pairs (in other words, a *value function*); in this way, the long-term value of a given state transition can be assessed, including the benefit of accessing "better" states later on by entering "worse" states in the interim.

By observing states, choosing actions, and assessing the efficacy of those actions over time, the agent eventually learns to choose actions such that the highest long-term reward is received. Some architectures for RL also use and/or develop a process model of the system to facilitate planning and learning.

There are a variety of ways in which the elements above can be computed, learned, and stored. The way in which they relate to one another in any given RL scheme also varies considerably. Some methods, for example, take the policy being followed into account during learning of the value function, whereas others do not. A good overview of the basic schemes and further discussion can be found in [17, 18], and discussion on NN implementations specifically can be found in [19].

Note that RL can be re-framed as a stochastic optimization problem in which one is searching the policy space directly. In particular, there is general interest in applying evolutionary computation methods to RL problems; however, the various merits of each approach have long been a subject of debate within the RL community. For some discussion, see [20].

### C. Artificial Neural Networks

*Artificial neural networks* (NNs) are particularly appealing tools for completing machine learning tasks and for creating intelligent agents. They are universal function approximators [21, 22] that are tailored specifically for a given task/computation. As such, they are highly flexible and are in principle able to operate effectively in many different situations and serve many different purposes.

In its simplest form, a NN consists of a collection of functions with weighted connections between them. These weighted connections can be adjusted ("trained") until a desired output behavior is achieved, typically through an automated optimization procedure. Elements of the network structure itself (for example, the number of nodes and layers), can also be adjusted as part of training.

NNs can be trained entirely from simulation data, entirely from measured data, or from a combination thereof. There are numerous architectures and training methods that are each suited to different kinds of problems (for an introduction, see [23]). For accessible overviews of basic concepts in NN-based control, see [24, 25].

### IV. NEURAL NETWORKS FOR PARTICLE ACCELERATORS

### A. Some Proposed Use Cases

In the context of modeling, diagnostic analysis, and control, there are many ways in which NNs can be used, and many of these are highly relevant to particle accelerators. Here, we propose some specific use cases of NNs that may be of interest to the particle accelerator community. The authors are actively pursuing several of these approaches.

1) *As an identified system model for control or simulation*

NNs are able to account for physical characteristics of systems which (a) have many interactions between a large number of parameters, (b) are not able to be realistically or completely modeled through analytic or standard simulation-based methods (due to practical considerations or limitations in the theoretical framework), and/or (c) vary significantly over time or involve behavior over multiple timescales. NN models can be deployed directly in model-based control routines. One area where NN-model based control is particularly appealing is in predictive control for accelerators, especially for subsystems where time delays and nonlinear behaviors are present.

In addition to simply creating a static model generated using previously gathered data, these models can be automatically updated over time to account for changing behavior due to drift or intentional changes in basic operating conditions. This can be done continuously, on a pre-set schedule, or when triggered by a specific event (such as a spike in modeling error). NN models can also be trained initially using data from an existing model or simulation first to speed up learning of un-modeled behavior once measured data is obtained.

NN models can also be used just as any other machine model might be (e.g. simulation for control design, simulations for experiments being planned for an existing machine, simulation of adjacent subsystems in the context of new components).

2) *As a fast stand-in for a computation whose speed usually limits real-time deployment*

Because the computation time for a trained NN is generally quite fast, they could be useful in cases where some known input-output relationship needs to be computed more rapidly for real-time deployment. For example, a NN model could be trained on high-fidelity physics-based simulation data and used as a fast, accurate stand-in for the full physics-based model.

Similarly, a NN could be trained by example to complete a computationally intensive control calculation or diagnostic procedure. For example, NNs can be trained to approximate the optimization procedure used in model predictive control to determine a future series of control actions [26, 27], thus enabling rapid computation of solutions during operation.

*3) As a way of doing fast, sophisticated diagnostic analysis and feedback*

NNs trained to do classifications, feature selection, and dimensional reduction could be used both for individual diagnostics and for collective higher-level control and machine protection.

For example, one could imagine a NN classifier that is used with an image-based accelerator diagnostic to sensitively identify when an undesirable beam instability is starting to develop, perhaps long before the effect is large enough for an operator to notice it.

NNs could also be used to provide bunch-to-bunch analysis and feedback, particularly if implemented in hardware or firmware. This avenue of study could prove to be particularly advantageous for accelerator systems with high repetition rates that require fast processing of diagnostic information and controller reactions.

*4) As a means of codifying and executing an existing policy that is not already codified*

For example, one could train a NN to mimic the observed behavior of an operator to automate some routine tuning task.

*5) As a means of improving an existing controller or optimization scheme*

NNs could be trained to mimic an existing controller or model and then improve upon its behavior through additional training during operation. This could be useful in cases where a traditional controller performs somewhat adequately but needs some small unknown adjustments.

NNs can also act as an adaptive "helper" function on top of another controller or model (e.g. as an adaptive addition to PID gains, or a nonlinear term on top of a linear model). As such, NN-based controllers can work in tandem with traditional control techniques to improve performance.

It is also useful to note that approaches using ML for self-tuning of optimization algorithm parameters are now appearing in the literature (e.g. see [15]). Such hybrid approaches to optimization could be readily put to use by the accelerator community in existing applications of online optimization.

*6) As part of a tool to learn and execute a new control policy*

NNs can be used to develop a control policy from scratch by examining the success of individual actions over time. In this way, instead of relying on an operator or an extensive control design procedure, the NN can discover the best way to interact with the machine. As with NN models, the solution that is finally deployed can be static or adaptive.

The main advantage of using NNs in this case (as opposed to other approaches used in the reinforcement learning) is that they enable continuous-valued functions to be estimated, enabling better generalization, and they can more efficiently represent the information that needs to be stored for large parameter spaces [18]. Their use also confers some added flexibility in the architecture: for example, one might have a joint actor/critic NN, as opposed to having two separate modules. Similarly, their ability to be trained as both models and as controllers can aid the process of learning in the latter capacity. For a particularly novel example of this, see [123], in which a pre-training step consisting of model learning is used to speed up learning of the value function in a NN-based RL control scheme.

### B. Historical Impediments and Recent Advances

*1) Recent Advances in Neural Networks and Their Deployment*

Neural networks have a somewhat tortured history, including a long series of boom-and-bust hype cycles. Early attempts at real-time control of complex systems with large parameter spaces using NNs were met with limited success, primarily due to issues with long computation times, a lack of sufficiently powerful architectures, and algorithmic instabilities. In the former case, early potential applications were limited by the computational speed of these techniques relative to the speed of the system dynamics to be controlled. Early attempts to apply these techniques in real-time to complicated problems were thus inherently limited, as only simple algorithms and structures could be investigated in the available time. These simplistic, early algorithms were also highly sensitive to small, arbitrary changes in input data, preventing the robust generation of solutions. In the interim, advances in the theoretical underpinnings of NNs have removed many of the previous impediments that once caused many would-be practitioners to abandon attempts to use NNs for control of complicated systems like particle accelerators.

Several major developments have since removed or mitigated many of these difficulties. First, improvements in computing technology over the past two decades have made successful implementation of more complicated NN structures and their training algorithms feasible in real-time applications. Such advancements have also greatly increased the speed of training in general, allowing much larger training data sets to be used in practice.

These improvements in computing have also allowed ever-larger data sets to be easily collected and stored. Along with this, the rise of the internet has enabled large stores of data to be accessed and shared by researchers world-wide, which both facilitates basic research and rigorous comparisons of algorithm performance.

In addition, many advances have been made in implementing NNs in hardware/firmware, such as FPGAs (e.g. see [28]) and neuromorphic chips. These include cutting-edge, high-end developments driven by large companies like IBM [124] as well as the production of more accessible commercial products (e.g. see General Vision's "BrainCard" [125]). A historical review on neuromorphic hardware can be found in [126], and an example of a comparative experimental study can be found in [127].

Next, beneficial co-developments in related fields such as stochastic optimization and computational learning theory have enabled more powerful learning algorithms to be developed. Similarly, advances in reinforcement learning, optimal control, adaptive control, and nonlinear control have facilitated successful deployment of NNs within these frameworks (e.g., see nonlinear model predictive control [121]). General theoretical research in the area of NN-based nonlinear control techniques also has been aimed at mitigating the issue of system stability in the context of control (see [29, 30, 31, 32]).

Critically, much theoretical work over the past two decades has been devoted to understanding the behavior NNs and developing more sophisticated architectures and associated training methodologies (see, for example, the more modern architectures described in [35]). Along with this, the growing body of experience with specific real-world applications has both informed theoretical developments and resulted in empirically-derived improvements in implementation and training procedures; the past five years have been particularly fruitful in this regard. Comprehensive reviews of these developments are given in [33, 34]. Some particularly important examples include:

—The introduction of selective data dropout techniques during training to reduce over-fitting [36];

—The introduction of an initial unsupervised learning stage in multi-layered, feed-forward NNs to capture data features and form progressively higher-level representations for subsequent layers prior to supervised learning [37, 38, 39, 40];

—Improvements in the training of recurrent NNs (RNNs) [41, 42], which contain recursive connections to enable representation of more complicated sequence-dependent dynamics (these recurrent connections also introduced additional mathematical difficulties for many gradient-based training algorithms [43, 44]);

—The development and advancement of long short term memory (LSTM) RNNs and their associated training techniques [45, 46, 47, 48, 49, 50], which can be used to more effectively represent long-term dependencies than many other architectures;

—GPU-accelerated training of convolutional NNs (CNNs) [51], which are designed to take more complete advantage of relationships within 2D input data (for example, they have substantially improved the state of the art in image-oriented tasks such as object recognition and speech recognition—where a 2D time-frequency representation is often used);

—The development of advanced neuroevolution methods such as NEAT and HyperNEAT [52].

Combined, these broad areas of advancement have enabled significantly more complicated problems to be effectively addressed both in theory and in practice.

*2) Previous Efforts to Apply Neural Networks to Particle Accelerator Control*

The idea of applying artificial intelligence and neural networks to particle accelerators is by no means a new one (e.g. see [53, 54, 55, 56, 57]). During an initial wave of interest during the early/mid-1990s, these efforts obtained mixed results. This quickly led to stagnation of efforts to apply AI to particle accelerators. A summary of work in this area up until 2008 is provided in [58].

As part of one notable dedicated effort in the mid-1990s, Vista Control Systems and University of New Mexico collaborated on the development of an AI-based beamline tuning prototype [59, 60, 61, 62]. Several studies also demonstrated the implementation of a distributed AI system for fault detection and management [63, 64]. Several recent (2012-2013) simulation-based studies propose some multi-agent designs for orbit control, beamline tuning, and mitigating the impact of sensor failures [65, 66, 67].

During the 1990s to mid-2000s, NNs in particular were investigated for orbit/trajectory control [68, 69, 70, 71, 72], with mixed results. Also during this time, a NN was successfully implemented to detect faulty beamline and diagnostic components [73]. In the early 1990s at Los Alamos, a NN-based PID tuner for a low level RF system was implemented [74]. Also at Los Alamos, several neural network schemes were used to control a negative ion source [75, 76, 77].

In work conducted at the Australian Synchrotron and the Linac Coherent Light Source, members of our group demonstrated the use of a combined NN and PI controller to compensate for jitter in the upstream klystron phase and voltage using downstream corrections, thus stabilizing electron beam energy and bunch length [78, 79]. In that control scheme, a NN was used to predict future beam parameter deviations so that an appropriate correction could be applied. In another study, a multi-agent NN tuning tool was used to optimize machine settings for reduced electron beam energy spread and increased transmission at the Australian Synchrotron Linac [80]. This optimization agent was then used in a control experiment at the FERMI@Elettra FEL to stabilize beam energy.

*3) Some Successes in Other Science and Engineering Disciplines*

Numerous examples of successes in using modern neural networks for various tasks are provided in [33] and [34]. Here we will specifically highlight a few examples that intuitively relate to the problems found in modeling, diagnostic analysis, and control of particle accelerators.

First, NN-based techniques to automatically process complicated measured scientific data have seen great success in recent years. For example, they have become useful tools in the analysis of astronomical data [e.g. see 81, 82, 83, 84, 85]. Some applications include image-based object classification from sky surveys and rejection of artifacts and interference in astronomical data. NNs have also proven useful in the analysis of data generated by high-energy particle physics experiments [86, 87, 88], which require detection and analysis of statistically improbable events from data sets with high backgrounds for event selection and particle identification.

Moving toward the application of NNs to real-time analysis of complex machines, an instructive example can be found in the recent literature surrounding fault prevention in tokamaks—a type of magnetic confinement device that is a promising candidate for thermonuclear fusion-based power production. In studies conducted within the last four years at the Joint European Torus, NNs have shown promise in the modeling of tokamak behavior [89, 90], in instability detection [91, 92], and in disruption prediction [93, 94]. Furthermore, real-time early detection of potentially problematic features such as hot spots or instabilities through the use of NN-based classification of video frames has been demonstrated experimentally [95, 96].

More generally, examples of NN applications in industrial control include the regulation of nonlinear chemical mixing processes [97, 98, 99], parameter/process optimization in manufacturing to achieve specific material properties [100] and greater product consistency [101], temperature control of variable-frequency systems with nonlinearities and time delays [102, 103], optimization for energy savings in the temperature control of buildings [104], self-tuning in traditional control schemes such as PID [105], and process optimization for reduced operating costs [106]. For a survey of reinforcement learning results in robotics, see [107]. Approaches that specifically pair advanced optimal control techniques with NNs have also been the subject of numerous successful experimental and simulation-based studies (see, for example, NN-based model predictive control [108, 109, 110, 111, 112, 113, 114]).

## V. An Example Application: Resonance Control at FAST

The system that is used to regulate the resonant frequency of the electron gun at the Fermilab Accelerator Science and Technology (FAST) facility was identified as a good initial candidate for the application of NN-based control methods. This was due to the large thermal time constants, long transport delays, and recursive behavior in the cooling system that collectively result in long settling times and large overshoots under PI control (these are described in parts *A.2* and *B* below, respectively).

The electron gun at FAST [115-117] is a 1½-cell copper RF photoinjector operating at 1.3 GHz in the $TM_{010,\pi}$ mode, and it is powered by a 5-MW klystron. It has a loaded Q of ~ 11,700, is water-cooled, and shows a measured 23-kHz shift in resonant frequency per °C change in cavity temperature. The gun is designed to produce 1-ms duration macropulses at a 1-Hz to 5-Hz repetition rate, with a bunch frequency of 3 MHz. The intended operational gradient is 40-45 MV/m, and the maximum gradient thus far achieved is 47.5 MV/m. For adequate phase stability, existing requirements state that the temperature of the water entering the gun should be regulated to within ± 0.02 °C [116]. At 40 MV/m and 5-Hz macropulse repetition rate, the expected average dissipated power in the gun is 15 kW.

### A. Water System Description and Control Challenges

A simplified schematic of the water system is given in Fig. 1. The two controllable variables are 1) the flow control valve setting (FCV) and 2) the heater power setting (HP). The T01 sensor reads the cold water supply temperature, the T02 sensor reads the temperature just after the mixing chamber, the TCAV sensor reads the cavity temperature, and the TOUT sensor reads the temperature of the water leaving the cavity. The TCAV sensor is located in the iris of the gun. Henceforth, we will generally refer to the abbreviated names only (e.g. "T01" indicates either "the T01 sensor" or "the T01 sensor reading").

TABLE I
TYPICAL TIME DELAYS BETWEEN SYSTEM ELEMENTS

| System Segment | Approximate Time [s] |
|---|---|
| Flow valve to T02 | 8-10 |
| Heater to T02 | 5-11 |
| T02 to TIN | 32 |
| TIN to TCAV | 19-23 |
| TOUT to T06 | 60 |
| TIN to cavity frequency | 16-18 |

Note that we have included transport delays as well as thermal responses; these are typical values seen under normal system operation.

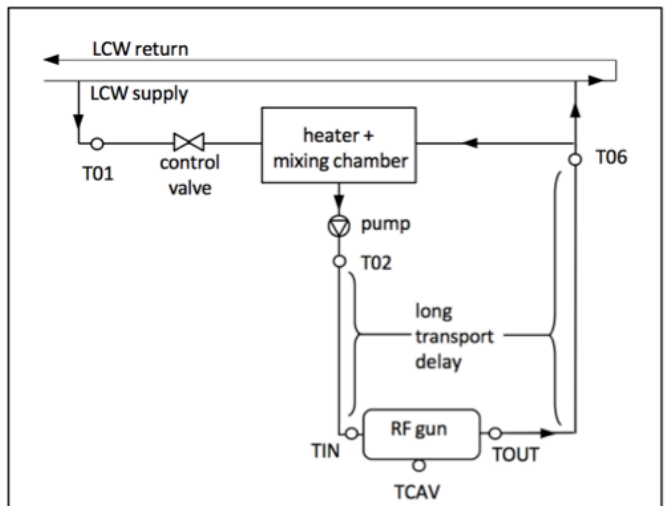

Figure 1: Layout of the water system and relevant instrumentation (not drawn to scale). T01, T02, TIN, TCAV, TOUT, and T06 are temperature sensors. The piping is not insulated. Note that for the regions marked "long transport delay," the piping traverses several parts of the building that at various times may have different ambient temperatures. The control valve and heater/mixing chamber are located outside of the radiation-shielding cave, whereas the gun is located inside of it.

*1) Instrumentation*

A description of the instrumentation is given in [118]. However, several important details regarding the resistance temperature detectors (RTDs) and the associated data acquisition process have changed.

The original analog-to-digital converter (ADC) hardware units were found to have unstable read-backs, likely due to differences in the exact versions of the MODBUS protocol used in those units and in the programmable logic computer (PLC) to which they were connected. They were subsequently replaced with RTD temperature transmitters made by *Laurel Electronics, Inc*. This hardware change resulted in lower-resolution readings in T01, T06, and T02 (0.1-°C resolution rather than 0.01-°C resolution). Noise in the readings typically results in a variation of ± 0.2 °C. Some of the data described in later sections were obtained under this configuration.

After several months, the conversion method for T02 was changed again in order to achieve higher resolution. It is now converted to a digital reading by a Fluke 8846A multimeter with a resolution of 0.01°C. The noise on the readings results in a variation of ± 0.02°C. This is the same ADC setup that is used for TIN, TCAV, and TOUT.

*2) Control Challenges*

For this particular system there are several control challenges:

—Due to water transport time and thermal time constants, long, variable time delays exist between various elements in the system. A selection of these is shown in Table I.

—Without compensation, any change in the temperature of the water exiting the gun (either due to a change in the amount of waste heat from the RF power or a change in the temperature of the water entering the gun) will circulate back into the mixing chamber and have a secondary impact on the cavity temperature. This results in a minutes-long, damped oscillation in the temperature of the water entering the gun. An example of such an oscillation (under no control, i.e. open loop) is shown in Fig. 2.

—There are fluctuations in the low conductivity water (LCW) supply temperature. While it is nominally regulated to within ± 0.5 °C, larger spikes do occur, especially during operation of other large heat sources in the wider system at FAST (e.g. the cryomodule high-level RF system is cooled by the same LCW supply that cools the RF gun).

—The pipes through which the water flows are not insulated and pass through several different areas of the building. Additionally, the closest ambient air temperature sensors, which read the south cave temperature and south hall temperature, show variations of several degrees day-to-day and more than 15 °C over longer durations. These two temperature readings are not always closely correlated. The relationships between T02, TIN, and TCAV vary measurably with these ambient temperatures, as does the relationship between TOUT and T06. Occasionally the steady state difference between TIN and T02 temporarily changes without a change in the ambient temperature readings being registered. Presumably, this is due to highly localized variation in temperature (e.g. cooling from a fan, heating from nearby equipment, etc.).

—Due to the TCAV sensor location and the cavity geometry, the temperature recorded there will be higher than the real bulk cavity temperature under RF power. Thus, for resonance control using operator-specified TCAV set points, it is important to note that the set point required to maintain the proper resonant frequency will increase with increasing average RF power. This is also a good reason to regulate the measured resonant frequency directly, rather than regulating the TCAV sensor read-back. An estimate of cavity temperature from TIN and TOUT could be used instead, but ultimately this is still a little circuitous relative to the end goal of keeping the gun at the desired resonant frequency.

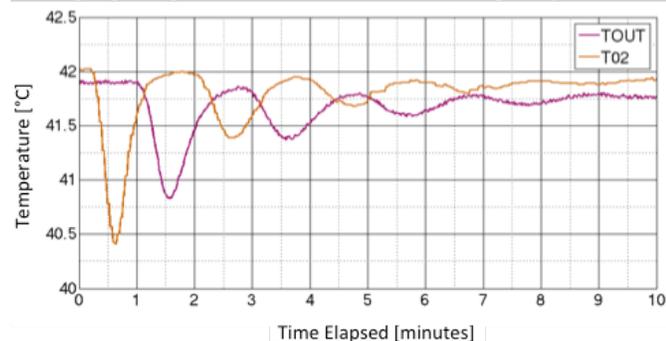

Figure 2: Oscillatory open-loop (uncontrolled) response in the water temperature at T02 due to mixing of the cold supply water and the water returning from the gun (TOUT). This oscillation was induced by reducing the heater power setting from 7 kW to 2.5 kW for 20 seconds, after which it was returned to 7 kW. Qualitatively, the response from an increase in the flow control valve is very similar (however, the system sensitivity to changes in the flow valve vs. the heater power setting are different).

### B. Description and Assessment of the Existing Feed-forward/PI Loop

Presently, the cavity temperature is regulated using a feed-forward/proportional-integral (PI) controller that was developed at Fermilab by P. Stabile (one of the authors). The feed-forward component is used to initialize the controller. It determines an appropriate flow control valve setting based on the RF power parameters and the expected cooling power of the water. It then continuously adjusts the valve setting such that a desired TCAV set point is reached, and the heater power level is kept at a constant setting. An older version of the controller and its performance is described in [118].

The response of the controller to a 1-°C step change in the set point under no RF power is shown in Fig. 3. The ~0.6-°C initial overshoot, the subsequent oscillations, and the long settling time are due to the combined effect of the long time delays and the recirculation of the water through the system. In the instance shown, the system takes ~23 minutes to reach a steady state. Note that this is without significant disturbances in the supply temperature (T01).

Typically, without disturbances from the RF power or T01, this controller regulates the TCAV temperature to within ± 0.03 °C of the set point during steady-state operation under RF power. The standard deviation of the TCAV temperature over 47,132 representative data points is 0.012 °C. This corresponds to the water temperature at TIN being kept within ± 0.04 °C of the mean temperature at steady state and a standard deviation of 0.013 °C.

While this is acceptable for long periods of steady-state operation, regulation using this controller becomes problematic under more dynamic conditions. For example, during RF turn-on, the overshoot results in reflected power often nearing and sometimes exceeding the threshold at which damage to ancillary components becomes a concern. To illustrate this, we examined 8 turn-on instances. Even with operator-mediated, gradual increase of RF power during normal operations at ~2.33 MW forward RF power, the controller initially overshoots the TCAV set point by an average of 0.19 °C (with a standard deviation of 0.02 °C). This results in a mean increase in reflected power of 70.6 kW over the steady-state value (with a standard deviation of 17.4 kW), culminating in a total mean reflected power of 103.0 kW. An administrative limit for reflected power at the RF window is set to 100 kW to avoid potential damage, and reaching it prompts the operator to turn off the gun. There is no self-excited loop mode implemented for this cavity to circumvent this issue during start-up.

Furthermore, in relying on manual adjustment of the TCAV set point for resonance control, operational time constraints combined with the long settling time make it less likely that the gun will be put at the desired resonant frequency consistently (and indeed, experience has borne this out).

This also reduces overall operational efficiency, as the low level RF system increases the forward RF power in response to the reduction in field caused by moving away from the desired resonant frequency. In light of this, it quickly becomes apparent that a controller capable of automatically making adjustments in the water system until the gun is operating at the proper resonant frequency (or some optimal distance off-resonance) is needed.

Overall, a significant improvement in the settling time, amount of overshoot, and disturbance rejection could be gained by adopting alternative control techniques. This would increase the operational efficiency of the gun by reducing the need to rely on the RF overhead to keep the cavity field constant, increase the total useful machine time by reducing the time spent waiting for the system to settle, and assist in the management of reflected power by more tightly regulating the cavity temperature under dynamic conditions.

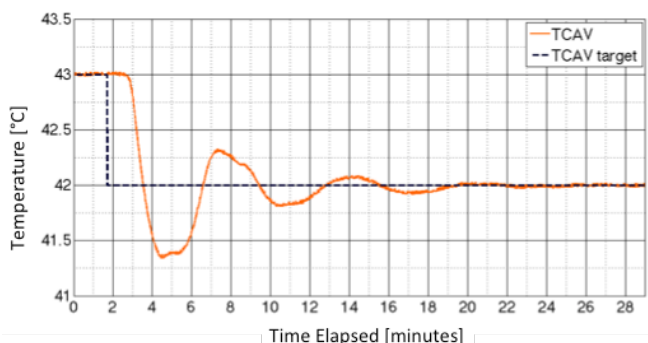

Figure 3: A 1-°C step change under the existing feed-forward/PI controller. Note that the oscillations are due to the time delays, thermal responses, and recurrent effect of the water system, not a poorly tuned set of PI gains.

E. *System Characterization*

The main goals for the characterization of this system were the following:
1) Accurately identify the transport delays;
2) Quantify the combined effects of TOUT, T01, FCV setting, and HP setting on T02 (and subsequently on TIN and TCAV);
3) Quantify the combined effects of TIN and RF power on TCAV;
4) Quantify the impact of ambient temperature on the temperature differences seen within the water system and the cavity (from TOUT to T06, from T02 to TIN, and from TIN to TCAV).

*1) Data Sets Obtained*

Figures 4-6 show a selection of the main data sets. Variables that relay redundant information or did not undergo significant changes are not shown.

The purpose of Set 1 (shown in Fig. 4) was to obtain data for many combinations of FCV and HP settings. The purpose of Set 2 (shown in Fig. 5) was to obtain data for several RF power settings, primarily targeting the relationship between TIN, RF heating, and TCAV. The purpose of Set 3 (shown in Fig. 6) was to obtain larger variations in the FCV and HP settings. Set 4 consists of changes in the temperature set point under PI control. Set 5 (shown in Fig. 7) consists of data gathered over several days during normal operation at a cavity gradient of 42 MV/m. The sensor replacements described earlier (section *A.1, Instrumentation*) occur after Sets 1 and Set 3. Several other smaller data sets were also examined.

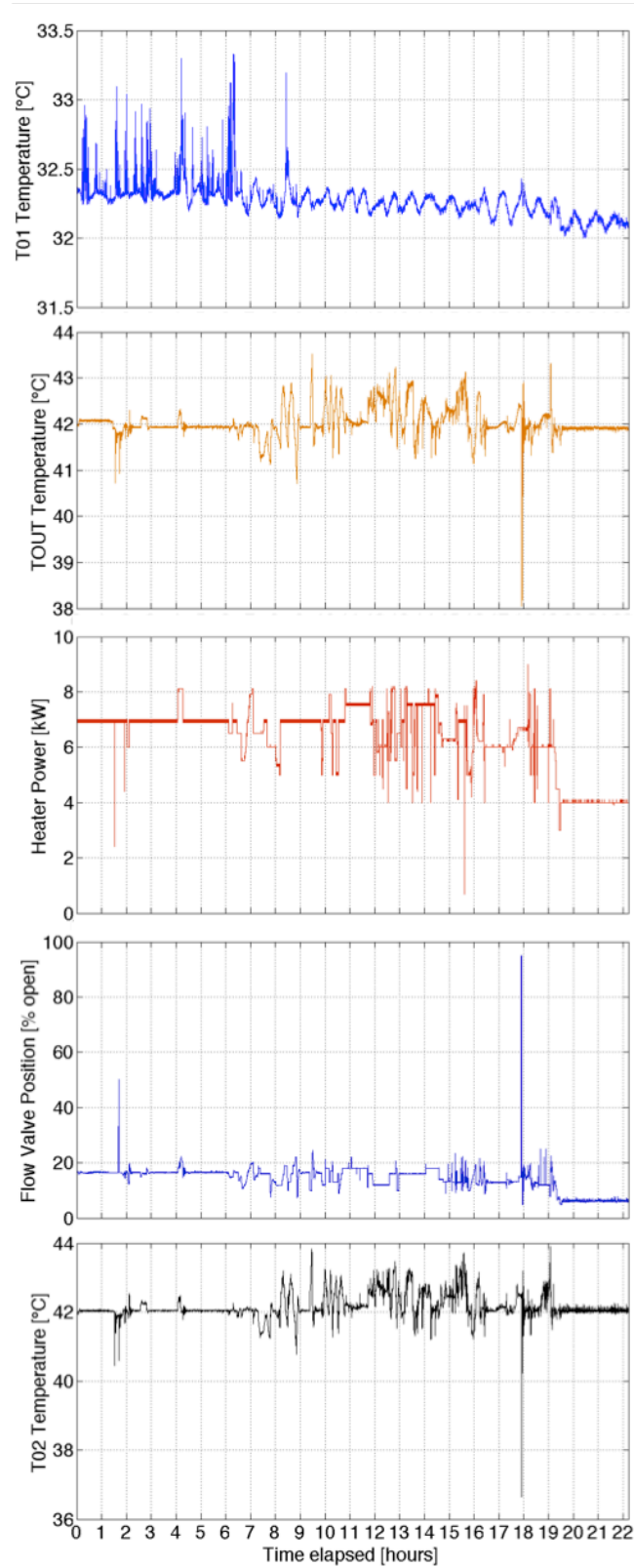

Figure 4: Data Set 1. This consists of water system data with the gun running at low average power (~81-kW forward RF power, 200-µs pulse duration, 1-Hz repetition rate). Changes to the FCV and HP settings were made in a pseudo-random fashion. The PI controller was not active.

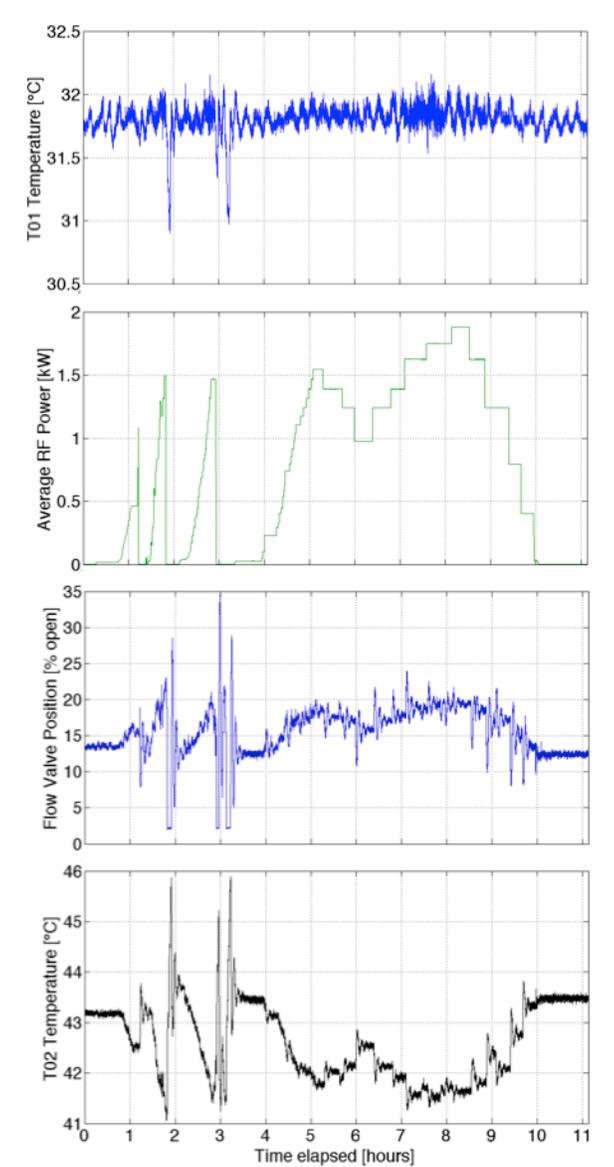

Figure 5: Data Set 2. This consists of system data under RF power to characterize the relationship between the temperature of the water entering the gun, the cavity temperature, and the RF power. For comparison with the other data sets shown here, T02, T01, FCV, and RF power are shown. Due to operator concerns about reflected power, the PI loop was enabled during this time. Note that in between Set 1 and Set 2, the ADC hardware for some of the sensors (T01, T06, T02) was changed, resulting in lower-resolution readings.

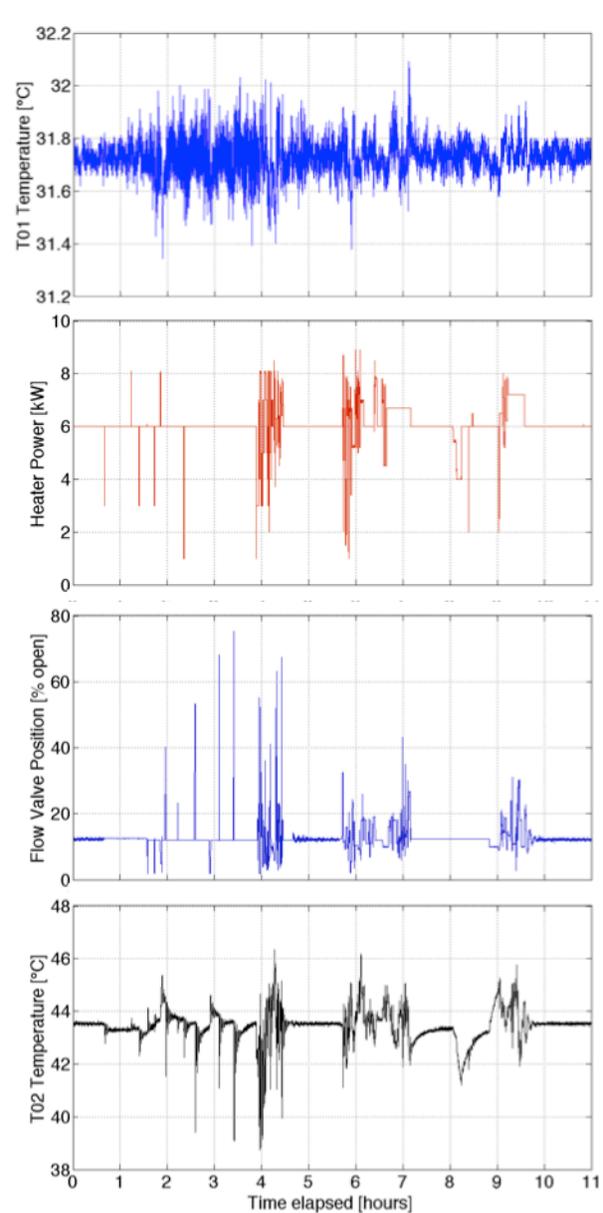

Figure 6: Data Set 3. This consists of higher-magnitude changes in the FCV and HP settings. The gun was off during this set, and the PI loop was not enabled.

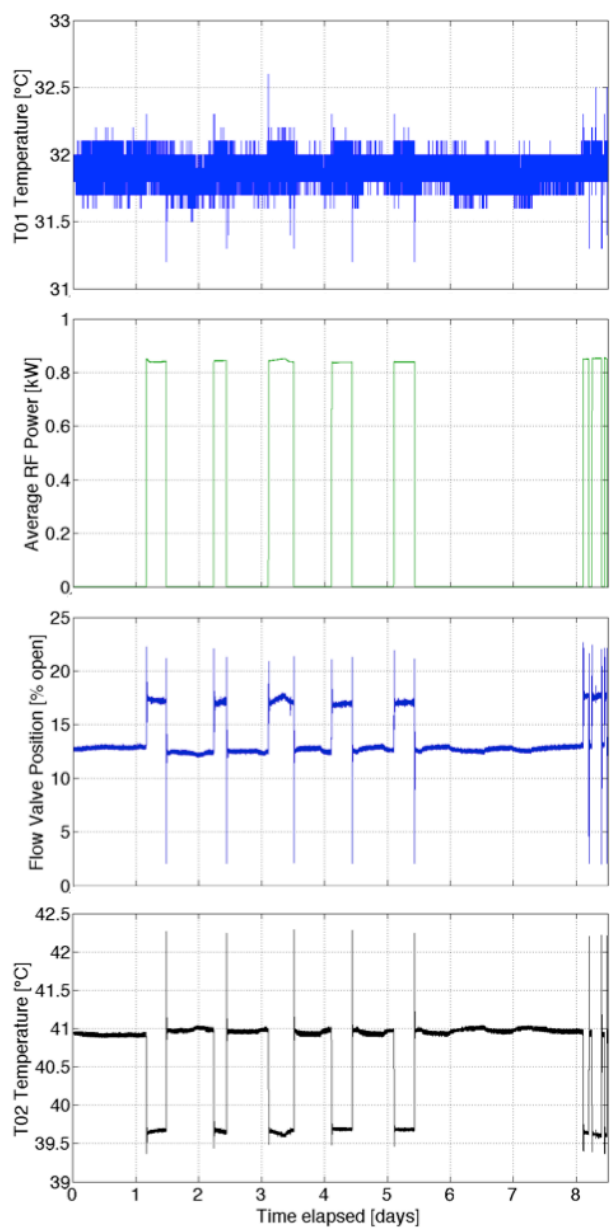

Figure 7: Data Set 5. This shows normal operations at 2.33-MW forward RF power. Note that, aside from the changes in the un-powered vs. powered state, smaller changes in the T02 temperature roughly inversely track changes in the ambient temperatures (shown in Fig. 8). This is because the relationship between T02 and TCAV slowly changes and the PI loop compensates for this by adjusting the FCV setting. Note also that operation of the gun does appear to be correlated with some variation in the LCW temperature (T01).

*2) Influence of Ambient Temperature*
The three primary areas where thermal losses could potentially impact the system are:
1)  The water in transit from T02 to TIN;
2)  The water in transit from TOUT to T06;
3)  Losses from the cavity surface to air or attached components.

Given the location of the ambient temperature sensors, their readings give us only a rough approximation of the air temperatures encountered by the bare pipes.  Figure 8 shows the variation in the south hall and south cave temperatures over the course of several days. Figure 9 shows the temperature difference between TCAV and TIN as a function of the difference between TCAV and the cave temperature.

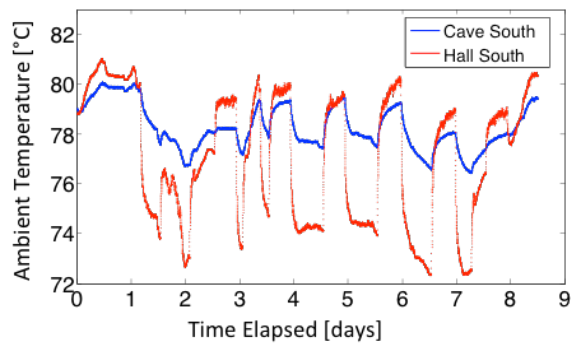

Figure 8: Variation in ambient air temperature over several days. A greater range of variation does occur over longer timescales.

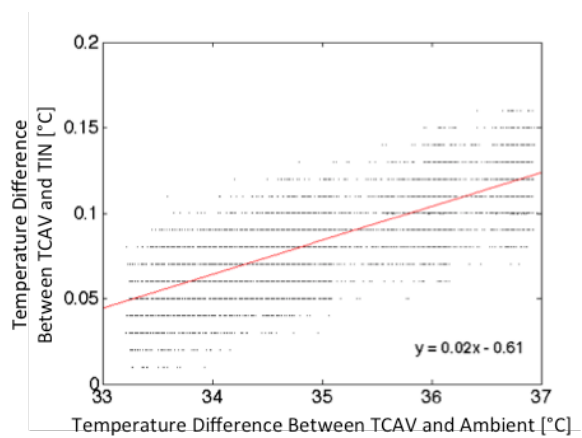

Figure 9: Temperature difference between TCAV and TIN, relative to the temperature difference between TCAV and the ambient air temperature reading.

The difference between T02 and TIN as a function of the estimated average ambient temperature shows a much less significant trend (the linear slope is 0.002), and the intercept changes slightly when the gun and its associated equipment is running, perhaps indicating that some additional local heating is raising the temperature at TIN.

*3) System Under Power*
Figure 10 shows the steady state difference between TOUT and TIN, and the steady state difference between TCAV and TIN, as a function of average RF power. This is useful for determining what set point is needed for TIN such that the desired cavity temperature reading under a given average RF power level is reached (note again that this is distinct from the real bulk cavity temperature). TOUT–TIN diverges significantly from TCAV–TIN. This demonstrates, in part, the effect that local heating from the RF has on the iris region of the gun where the TCAV sensor is located. At steady state, the cooling power given by $P_{cool} = (T_{OUT} - T_{IN})$ x *(Flow rate [GPM]/ water cooling capacity [GPM-°C/kW])* is balanced with the power input to the cavity ($P_{cool} = P_{IN} \approx P_{RFavg}$). This relationship is shown in the dashed line on Figure 10.

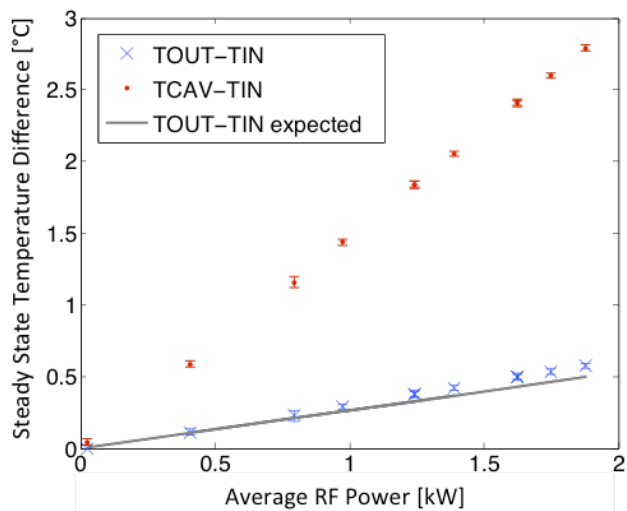

Figure 10: Difference between TIN and TCAV sensor readings, as well as TOUT and TIN sensor readings, as a function of RF power at steady state. The data were recorded at relatively constant cave temperature and a constant flow of water to the gun at 14.5 GPM.

### F. Neural Network Modeling

In modeling the system, the primary aim was to investigate several model structures and use these to inform the controller design. While elements of the system can be modeled analytically or with other data-driven techniques, by using NNs we are also setting the stage for NN-centric controllers.

The main variables examined were 1) the choice of model inputs, 2) the combination of training data to use, and 3) the NN architecture. In addition to training and testing with previously-gathered data, a few studies in online updating during operation were conducted. The following general model structures were derived from training data and assessed:

1) A model to predict T02 from TOUT, T01, FCV, and HP readings;
2) A model to predict TCAV from TOUT, T01, FCV, HP, and RF power readings;
3) A model to predict TCAV from TIN and RF power readings, and a similar model that uses T02 instead of TIN;
4) A model to predict T02 that also includes the south cave and south hall temperature readings as inputs.

Note that these models were developed in parallel with both control development and analysis of the impact of ambient temperature. In retrospect, it is likely that ambient temperature is only needed for the TCAV model. Because the benchmark controller (described later) primarily relies on a T02 model, a TCAV model with ambient temperature included as an input has not yet been constructed, but likely will be in the future.

#### 1) Data Preprocessing

The mean values were subtracted from the data, and the data were scaled to a range of ± 0.5. The 2.44-°C offset in T02 for Sets 1-4 relative to Set 5 was adjusted for by subtracting it from the measured values. In addition, a zero-phase digital filter was applied to T02 readings for data sets containing the noisier, lower-resolution data. The remaining data were not filtered. This produced more consistent results than either excluding any filtering procedure or filtering all of the data and using incremental filtering during online testing.

#### 2) Training Procedure

The training procedure described applies to all trials for ruling out different model input-output structures and NN architectures. The training data were used to teach the NN the proper input-output relationship via supervised learning. In this case, the Broydon-Fletcher-Goldfarb-Shanno (BFGS) algorithm, a popular quasi-Newton optimization method for unconstrained nonlinear problems, was used to find the optimal weights and biases of the network. During training, every other input-output pair was used for calculating weight updates, while the remaining data were used for validation (i.e. assessment during the training process). Validation data are used to ensure that the solution is generalizable (i.e. the model is not over-fitted to the training data and the proper relationships have been learned). The predictive performance on the testing data was used to assess the models, both in one-step-ahead prediction and in larger prediction horizons (20-200 steps ahead). To help ensure that the particular training algorithm used was not skewing the results significantly, results were compared with those obtained using the Levenberg-Marquardt algorithm (LMA).

For each candidate model architecture and set of inputs, multiple individual networks were trained and subsequently tested on the remaining data sets. This helps to ensure that the network architecture and configuration of the training data itself is responsible for the performance, rather than a single exceptionally poor or good solution.

It was found that there were various merits to training on Set 1 (which had higher-resolution T02 readings and contained a variety of flow valve and heater setting combinations, but also consisted of relatively small adjustments) and on Set 3 (which had fewer, but higher-magnitude adjustments and lower-resolution T02 readings). The best performance was obtained when the networks were trained first on Set 1 and then subsequently trained on a portion of Set 3 (with the rest of Set 3 being reserved for testing along with the remaining data sets). This is likely due to the fact that Set 3 had higher-magnitude changes than Set 1, but Set 1 had lower noise on the readings and thus the initial learned relationship had higher sensitivity to small changes in input parameters. Set 1 also consisted of many more individual training examples than Set 3, so training on the latter likely does not impact the final solution as much as the former. For the TCAV models, additional training was also conducted with a selection of data from Set 2 under RF power.

#### 3) Neural Network Structures and Model Inputs

Starting with a simple approach, a feed-forward architecture was adopted. Because the time delays can vary, a short series of previous values is given for each input. For long transport delays, only relevant values are provided (i.e. dead time is removed by introducing a delay in the inputs).

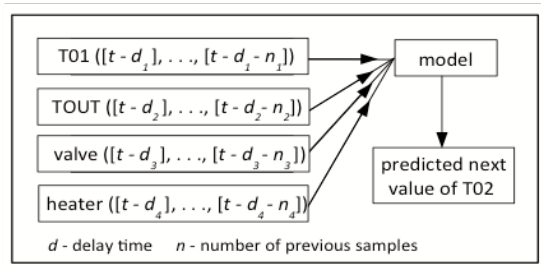

Figure 11: Simplified illustration of the T02 NN model inputs and outputs. A series of *n* previously measured samples are provided. Initial previous inputs are delayed by an amount *d* governed by system dead times.

For the T02 models, the best performance was obtained when the NN was provided with 15 prior seconds of FCV, HP, and T01 readings, and 30 seconds of prior TOUT data. The TOUT data inputs are delayed by 50 seconds and interspersed at a varied interval such that the 60-70 second range is represented more heavily (the intervals were determined experimentally). It was also found that the T02 models generally performed better when excluding the ambient temperatures. In addition, better solutions for long-term future predictions were obtained when excluding previous values of T02 as an input (thus forcing the NN to really learn the relationships between T02 and the other input variables, rather than relying on autocorrelation).

For the T02 models, a variety of configurations were examined (including variations in the number of layers, the number of nodes in each layer, and the type of activation functions used). Out of these, the best performing models used two hidden layers, with 20 hidden nodes in the first layer and 5 hidden nodes in the second layer. The nodes in the hidden layers use an approximate hyperbolic tangent sigmoidal activation function, given by $f(x) = \frac{2}{(1+e^{-2x})} - 1$.

The performance of the best models is given in Table II. The T02 models were the most extensively trained and vetted, and the final T02 model performs well across all data sets. The best-performing full TCAV model is very close in structure to the T02 model. Without power, it performs better than the T02 model (even without ambient temperature included). This may be due to the sets of lower-resolution T02 measurements included in training the latter, or due to the fact that TCAV is less susceptible to oscillations and noise in the water system because of its large thermal mass. In testing the TCAV models under RF power, there were steady-state offsets in the predicted output. Additional training under a greater variety of RF power levels would be needed to improve this.

TABLE II
AVERAGE PERFORMANCE OF SELECTED NN MODEL DESIGNS

| NN Model | Mean Absolute Error | STD of Error | Max. Error |
| --- | --- | --- | --- |
| T02 Sigmoid | 0.018 | 0.037 | 1.049 |
| T02 Sigmoid w/Ambient Temp. | 0.022 | 0.043 | 1.317 |
| T02 Linear | 0.058 | 0.266 | 2.915 |
| TCAV Sigmoid (tested without power) | 0.011 | 0.012 | 0.131 |
| TCAV Sigmoid (tested with power) | 0.259 | 0.287 | 1.390 |

Table II shows the performance of several NN model designs. For the best-performing network out of each model category, the average absolute error across all prediction instances in all of the data sets is reported, along with the standard deviation. The maximum error out of all data sets is also reported. For TCAV, the performance with and without RF power is reported separately. "Linear" and "Sigmoid" denote the activation function types.

### G. Control Over the Water System

Eventually, the aim is to have a NN controller that adjusts the FCV and HP settings such that the desired resonant frequency or some optimal amount of detuning is achieved. An additional aim is to control the rate at which RF power is brought up to its operational level during turn-on.

Establishing satisfactory control of the water temperature at the cavity entrance is the first step toward ensuring the gun is kept at the proper resonant frequency. Because the long transport delays and recursion in the water system are a major challenge for the feed-forward/PI controller, it makes sense to address this problem individually before moving on to a complete controller. Furthermore, while the cavity temperature (as reported by the TCAV sensor) is just an intermediate variable when considering the final goal, framing an initial controller around the water entering the gun and TCAV enables direct comparison with the existing feed-forward/PI loop.

As such, we aimed to design a modular controller that could be altered with ease to fit either TCAV-oriented regulation or resonant frequency-oriented regulation. The base controller regulates the temperature of the water entering the gun by modulating the FCV and HP settings. This controller can then be nested within another control loop that determines what the water temperature needs to be in order to either a) directly minimize the detuning or b) achieve an operator-specified cavity temperature set point.

### H. Benchmark Controller for the RF Gun

First, we developed a simple benchmark controller for the water system. This serves two purposes: 1) it provides a more suitable benchmark than feed-forward/PID by which to judge the performance of more advanced NN controllers, and 2) it provides an initial policy that a NN can be trained to mimic and improve upon.

Because of the long time constants, the effect of the water returning from the gun, and the presence of two controllable variables, a model predictive control (MPC) [119,120,121] scheme is appealing. In MPC, a system model and an optimization algorithm are used in conjunction to determine an optimal sequence of future controller actions such that the target output is reached within some future time horizon, subject to the satisfaction of any defined constraints. Such a scheme is useful for compensation of delayed system behavior. In addition, if a series of future set points is known in advance, the controller can act anticipatively. Figures 12 and 13 illustrate the basic concept of MPC. Because MPC relies on repeatedly computing an optimal future trajectory for a series of future time steps, there is a substantial tradeoff between model complexity and the ability to obtain a good solution within the control interval.

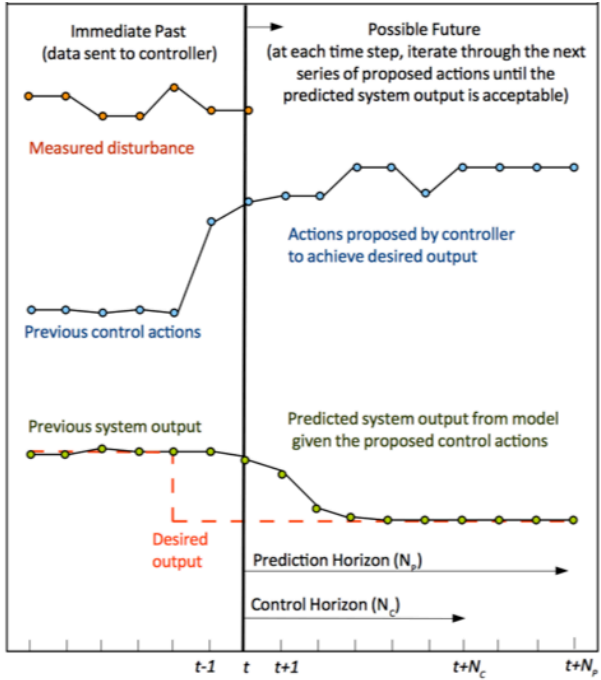

Figure 12: The basic concept of model predictive control.

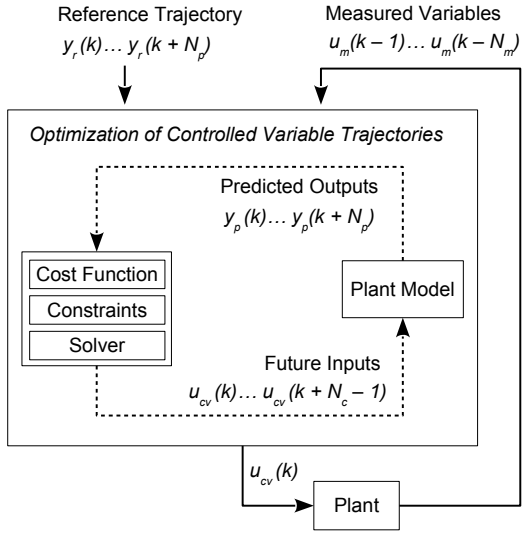

Figure 13: The basic elements of a model predictive control scheme. $N_p$ is the prediction horizon, $N_m$ is the number of previous measured values used for modeling, k is the present time step, $N_c$ is the control horizon, $u_{cv}$ are the controlled variables, $u_m$ are measured variables, and $y_p$ is the predicted plant output.

*1) Controller Structure*

T02 was chosen as the variable to control for the benchmark system. The basic structure of the benchmark MPC is shown in Figure 14. First, the operator provides a TCAV set point, which is then translated into an approximate T02 set point by exploiting the relationship between average RF power, T02, and TCAV. The MPC then manipulates the FCV and HP settings such that the desired T02 trajectory is obtained.

By monitoring temperature changes in the water leaving the gun, the controller can compensate for them before they reach T02. Monitoring TOUT provides plenty of time for actuation of the heater to take effect. By also monitoring T01 and using this as a model input, adjustments can be made to compensate for fluctuations in the LCW supply temperature. One critical weakness of this design is that the ambient temperature can affect the relationship between T06 and TOUT, and at present this is unaccounted for (e.g. by model updating or an adaptive offset on the TOUT input).

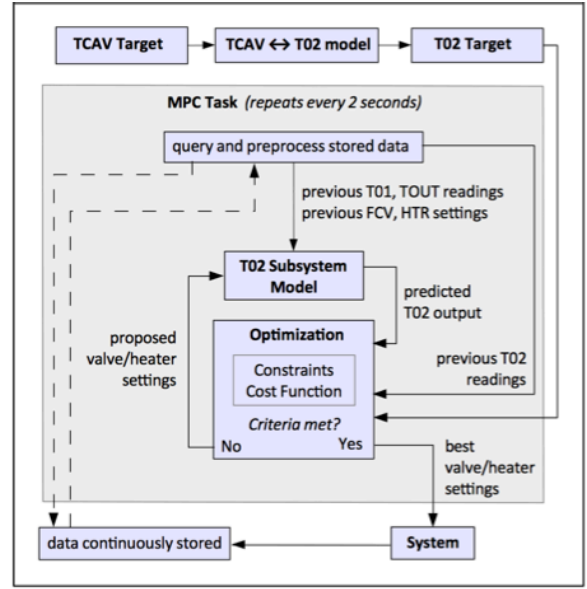

Figure 14: Conceptual structure of the benchmark MPC for temperature control of the RF gun.

For the optimization of future controller actions, the cost function is defined by the weighted rate of controllable variable changes, the weighted discrepancy between the present predicted output trajectory and the output reference trajectory, the weighted discrepancy between the desired trajectory of controllable variables and their present trajectory (if applicable), and the weighted degree to which any constraints are violated. Each of these terms is evaluated over the entire prediction horizon. If desired, small constraint violations are allowed via constraint softening. The general formulation for the cost function at time-step $k$ for one output variable is given by:

$$\sum_{j=1}^{n_{cv}} \sum_{i=0}^{N_p-1} \{w_{1,j}[u_j(k+i) - u_j(k+i-1)]\}^2 \\ + \sum_{j=1}^{n_{cv}} \sum_{i=0}^{N_p-1} \{w_{2,j}[u_j(k+i) - u_{j,ref}(k+i)]\}^2 \quad (1) \\ + \sum_{i=1}^{N_p} \{w_3[y_r(k+i) - y_p(k+i)]\}^2 + w_4 \, b \, ,$$

where $w_{1,j}$ is the weight for rates of change in the $j^{th}$ controllable variable, $w_{2,j}$ is the weight for the $j^{th}$ controllable variable target trajectory, $w_3$ is the weight for output variable target trajectory, $w_4$ is a penalty weight for constraint softening, $N_p$ is the prediction horizon, $i$ is a future time step, $y_r$ is the reference trajectory, $y_p$ is the predicted output, $k$ is the present control step, $u_j$ is $j^{th}$ the controllable variable value, $u_{j,ref}$ is the reference trajectory for the $j^{th}$ controllable variable, $b$ is a measure of constraint violation, and $n_{cv}$ is the number of controllable variables. The constraints are given by:

$$y_{min} - ba_{y_{min}} \leq y(k+i) \leq y_{max} + ba_{y_{max}}$$
$$u_{j,min} - ba_{u_{j,min}} \leq u(k+i-1) \leq u_{j,max} + ba_{u_{j,min}} \quad (2)$$
$$\Delta u_{j,min} - ba_{\Delta u_{j,min}} \leq \Delta u_j(k+i-1) \leq \Delta u_{j,max} + ba_{\Delta u_{j,max}}$$

where $\Delta u_j$ is the change in the $j^{th}$ controllable variable, $a_{u_{j,min}}$ and $a_{u_{j,max}}$ are variables for constraint softening for the $j^{th}$ controllable variable limits, $a_{\Delta u_{j,min}}$ and $a_{\Delta u_{j,max}}$ are variables for constraint softening for the $j^{th}$ controllable variable movement, $a_{y_{min}}$ and $a_{y_{max}}$ are variables for constraint softening for the output variable limits, and $b$ is a slack variable.

*2) Implementation*

For the benchmark MPC, we were only concerned with getting a rough idea of how well a simple MPC system might perform compared with the existing feed-forward/PI loop. To this end, we linearized the NN model around the present operating point at each time step and used the sequential quadratic programming solver QPKWIK [122].

In the benchmark controller, all constraints were hard (i.e. $a_{u_{j,min}} = a_{u_{j,max}} = a_{\Delta u_{j,min}} = a_{\Delta u_{j,max}} = a_{y_{min}} = a_{y_{max}} = 0$), and y was unconstrained. Because there is no constraint softening, $w_4$ was set to 0. Finally, there is no specifically desired controllable variable trajectory; as such, $w_2$ was also set to 0. Through a combination of simulation and testing on the gun, a set of the remaining MPC parameters that achieve reasonably good performance were obtained. These parameters are given in Table III.

A rudimentary NN model was used to translate between the TCAV set point and the T02 set point. One could instead use the simple steady-state relationship, but we wanted to capture the dynamic response as well.

TABLE III
BENCHMARK MPC PARAMETERS

| Parameter | Value | Units |
|---|---|---|
| Valve max rate | 10 | % open/sec |
| Valve upper limit | 70 | % open |
| Valve lower limit | 2 | % open |
| Heater max rate | 4 | kW/sec |
| Heater upper limit | 9 | kW |
| Heater lower limit | 1 | kW |
| Prediction horizon | 100 | s |
| Control horizon | 20 | s |
| Control interval | 2 | s |
| Valve rate weight | 0.4 | - |
| Heater rate weight | 0.5 | - |
| T02 output weight | 0.3 | - |

*3) Performance*

Figure 15 shows the performance of the benchmark MPC for a 1-°C step change in the TCAV set point. The settling time is ~5x faster than that of the pre-existing feed-forward/PI loop, and there is virtually no overshoot. After the step command for the cavity is issued, the MPC brings T02 to within ± 0.02 °C of its respective set point in about 3 minutes. Correspondingly, TCAV is brought to within ± 0.02 °C of its set point in about 5 minutes. While the transient behavior in this instance is clearly an improvement over the transient behavior of the PI loop, additional data is needed to fully characterize both the transient and steady state performance.

Note that the scales in Fig. 15 are smaller than those shown for the feed-forward/PI controller in Fig. 2 (1.5-°C vertical extent in the former vs. 2.5-°C vertical extent in the latter, and 10-minutes extent in the former vs. 30-minutes extent in the latter). As with the results shown in Fig. 2, no RF power is going to the gun.

Figure 16 shows the measured FCV and HP actions, and Figure 17 shows the requested FCV and HP actions. We see an initial adjustment (the valve opens and the heater power decreases), followed by an adjustment in the opposite direction to compensate for the lower temperature of the water exiting the gun.

Overall, despite this being just a simple benchmark test, the performance achieved was a substantial improvement over that obtained with the pre-existing feed-forward/PI controller.

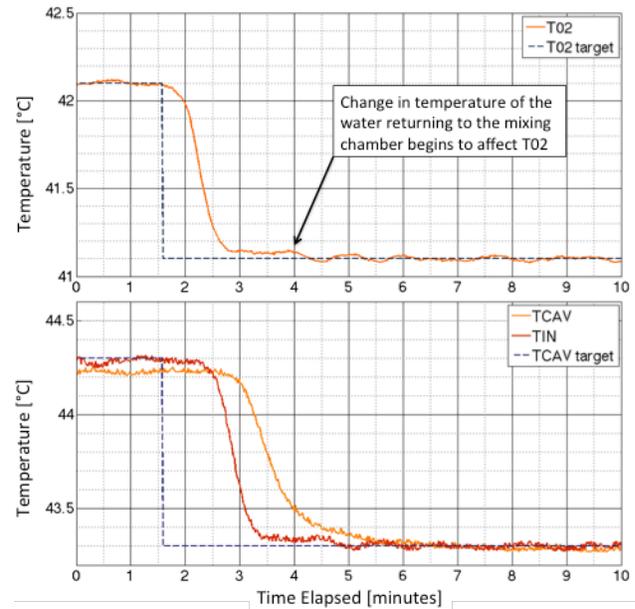

Figure 15: A 1-°C step change in TCAV under the benchmark MPC. Note that the scales are smaller than those of Fig. 2. These data were recorded as part of a series of steps in the TCAV set point. Note that this is not a perfect 1-°C step, as there is an offset between the original TCAV set point and the final value it obtained in the prior to step.

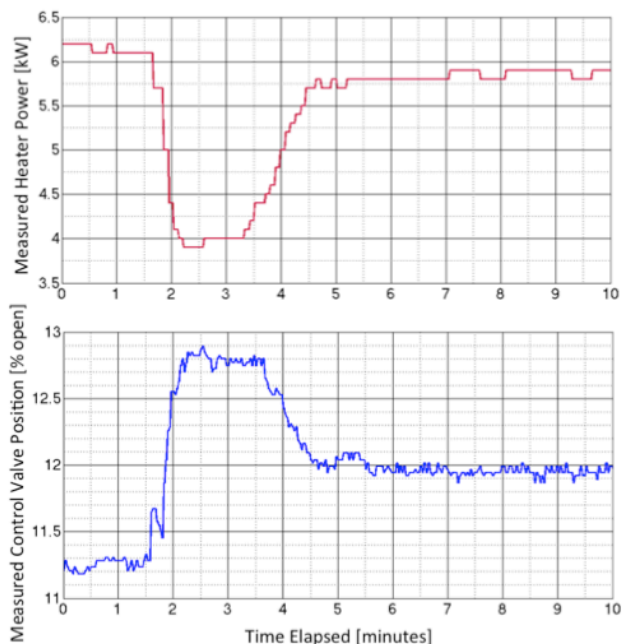

Figure 16: Measured flow control valve and heater power actions.

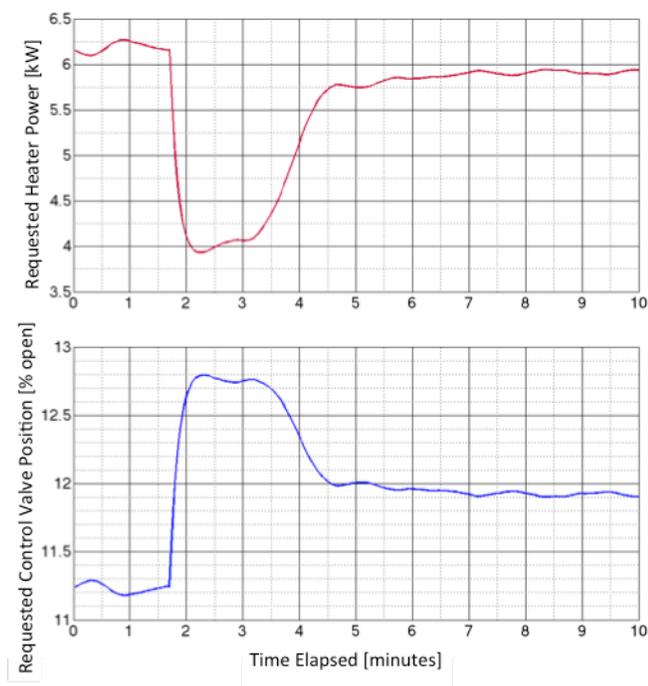

Figure 17: Requested flow control valve and heater power actions.

*4) Limitations and Potential Improvements*

First, improvement could be made to the timing of control actions. The small oscillations in T02 that start around the 4-minute mark are the result of imperfect timing in the compensative actions for the recirculating water. The oscillations were replicated in simulation by introducing a mis-match in the timing of TOUT in the plant relative to that in the model used for the MPC controller. During training, the time delay between when the controller issues a command, when the command is received by ACNET (Fermilab's main control system), and when the actuators in the hardware respond were not accounted for.

Another limitation of the controller is that it used previous requested FCV and HP settings in the model, rather than using measured values. Given how much the requested values deviated from the measured values, the performance of the controller could likely be improved by using the measured values instead.

The controller also needs to be tested over the RF power range of the cavity. The T02 model performs well under powered conditions, and thus in principle the MPC should be able to compensate for temperature changes in the water exiting the gun associated with RF power adjustments. However, the component that converts the TCAV set point to a T02 set point needs to be more carefully designed before this is implemented for regular use. Steady state offsets in the modeling could be accounted for by adding slow feedback to the component that translates between TCAV and T02 under RF power. Alternatively, additional training data under a wider variety of RF power levels could be obtained, and ambient temperature could also be included as an input. Because TCAV does have a slow thermal response, an extension of this is to use a second MPC to determine a desirable T02 trajectory (rather than a single set point) or simply lump the whole system together in one MPC.

*I. Future Work and Extensions To Pure NN Control*

The future work at FAST falls into two categories:
1) Expansion of and improvement upon the benchmark MPC
2) NN-centric schemes that build on (1)

The primary interest in (1) is that it can facilitate the learning of good initial controllers in (2). A number of possible improvements to the MPC are discussed in section *H.4*, but ultimately for our purposes it just needs to be good enough to provide a starting point for training. For control of the water entering the gun, the benchmark controller is good enough. We are training a NN to mimic the benchmark MPC behavior. This initial policy will subsequently be improved through additional training during simulated interaction with the machine. Finally, it will be experimentally tested. This module can then be extended to direct resonance control using another MPC or a reinforcement learning component.

Extending that approach further, we are creating a separate MPC unit that determines trajectories for T02 and RF power such that the requested operational RF power is reached without significant increases in reflected power. This would use resonant frequency (as measured by detuning) as the reference parameter. We will also examine using a NN-based reinforcement learning controller directly from scratch for this application (i.e. one that operates without the benefit of learning an initial policy from the MPC).

At present, we are also working on resonance control for an RFQ that will be used in the PXIE accelerator at Fermilab.

VI. CONCLUSION

There is a clear need for the development and validation of reliable, adaptive control techniques for complex problems in particle accelerators. Achieving this objective becomes particularly challenging as these systems achieve higher energies and intensities, are required to operate with ever-more stringent tolerances on beam parameters, and are needed for a rapidly growing range of applications with highly varied requirements. Incorporating artificial intelligence and machine learning techniques into particle accelerator control systems could greatly assist the community in its effort to meet these challenges.

The work at FAST provides just one example of how more advanced control methods that include learned models and planning can provide advantages over primarily reactive control paradigms. In that example, the presence of long, variable time delays and multiple controllable parameters resulted in a challenging control problem that was not handled effectively by the existing feed-forward/PI controller. A relatively simple neural network-based model predictive controller was able to substantially improve the control over the system, resulting in a ~5x shorter setting time and virtually no overshoot in the target parameter. Coupled with the previous work conducted by our group at LCLS, the FERMI@Elettra FEL, and Australian Synchrotron, it is apparent that modern AI- and machine learning-based control techniques can be put to highly effective use in particle accelerators.

Within artificial intelligence and machine learning, we highlighted neural networks specifically. Neural networks are highly flexible tools that could be used in many ways to improve particle accelerator performance. They can be used to learn system relationships (e.g. dynamic models), to do perform rapid computations, and to act as controllers. Myriad advances over the past decade have greatly improved the practicality of these techniques.

Furthermore, accelerators are useful test-beds for neural network-based control. By focusing simultaneously on incremental development and experimental testing, actual problems encountered in accelerator control can guide algorithmic development, resulting in a suite of new techniques that are uniquely well-suited to the field's operational challenges.

We plan to continue our research in this area to develop advanced controllers based on artificial intelligence and machine learning for a wide variety of control problems found in particle accelerators.

ACKNOWLEDGEMENTS

We would like to thank Elvin Harms for generously providing machine time at FAST, and we would also like to thank Jinhao Ruan, James Santucci, and Darren Crawford for their assistance during these studies. We are also very appreciative of the rest of the FAST team for their continued support and interest in this work. Many thanks also go to Denise Finstrom and James Patrick, who provided extensive assistance with the interface between our control code and ACNET. James Smedinghoff, Dennis Nicklaus, and Tim Zingelman also were of great help in exploring our options for interfacing with ACNET. Finally, we would also like to thank Curtis Baffes, Maurice Ball, Kermit Carlson, Jerzy Czajkowski, Nathan Eddy, Paul Kasley, and Thomas Zuchnik for their fruitful discussions with us about various system hardware components.